\pdfoutput=1  
\documentclass[11pt]{article}
\usepackage{arxiv}                  

\usepackage[T1]{fontenc}
\usepackage{amssymb}
\usepackage{amsmath}
\usepackage{multirow}
\usepackage{array}
\usepackage{makecell}
\usepackage{float}
\usepackage{placeins}
\usepackage{caption}
\usepackage{graphicx}
\usepackage{subcaption}
\usepackage{siunitx}
\usepackage[hidelinks]{hyperref}

\graphicspath{{figs/}}

\renewcommand{\shorttitle}{VGG16-MCA UNet for Whole-Tumor FLAIR Segmentation}
\renewcommand{\headeright}{A Preprint}
\renewcommand{\undertitle}{A Preprint}

\title{VGG16-MCA UNet: Whole-Tumor Segmentation\\ in 2D FLAIR MRI with Decoder-Side Channel Attention}

\author{%
  \textbf{Shubham Gajjar}\textsuperscript{\,1,\,*}\quad
  \textbf{Deep Joshi}\textsuperscript{\,2,\,\dag}\quad
  \textbf{Avi Poptani}\textsuperscript{\,2,\,\dag}\quad
  \textbf{Vishal Barot}\textsuperscript{\,2,\,\ddag}\\[0.8em]
  \textsuperscript{1}Northeastern University, Portland, ME, USA\\[0.3em]
  \textsuperscript{2}LDRP Institute of Technology and Research, Kadi Sarva Vishwavidyalaya,\\
  Sarva Vidyalaya Kelavani Mandal, Gandhinagar, Gujarat, INDIA\\[0.8em]
  \begin{minipage}{0.92\textwidth}
    \centering\small
    \textsuperscript{*}Corresponding author. Email: \texttt{gajjar.shu@northeastern.edu}\\
    \textsuperscript{\dag}Research Scholar, Department of Computer Engineering, LDRP Institute of
    Technology and Research, Kadi Sarva Vishwavidyalaya\\
    \textsuperscript{\ddag}Assistant Professor, Department of Computer Engineering, LDRP Institute of
    Technology and Research, Kadi Sarva Vishwavidyalaya
  \end{minipage}%
}
\date{}

\begin{document}
\emergencystretch=\maxdimen
\maketitle

\begin{abstract}
Automated brain tumor segmentation supports diagnosis, treatment planning, and monitoring of disease progression, but building models that generalize across heterogeneous tumors and limited annotated data remains difficult. We present VGG16-MCA UNet, a hybrid architecture pairing an ImageNet-pretrained VGG16 encoder with a decoder in which a Multi-Channel Attention (MCA) module recalibrates features after each skip-connection fusion, trained with the Focal Tversky loss to counter severe foreground-background imbalance. We evaluate the model as a 2D, FLAIR-only, whole-tumor segmenter on tumor-positive slices from two public datasets: the BraTS 2020 benchmark and the LGG MRI Segmentation dataset. Using 5-fold cross-validation and a single network formed by averaging the weights of the five fold models, the method attains an aggregate pixel-level Dice (F1) of 95.10\% on our held-out BraTS 2020 split and 88.32\% on LGG. These scores are computed over all test pixels pooled into a single confusion matrix rather than averaged per case, and are therefore not directly comparable to the per-case mean Dice used in the BraTS challenge protocol. All partitions were drawn over individual slices rather than over patients, so every patient contributes slices to both training and test; the figures above therefore measure interpolation within known patients and should be read as an upper bound rather than as generalization to new ones. The model segments a 256$\times$256 slice in 66.32 ms on a single 6 GB NVIDIA RTX 2060, approximately 8 ms more than an equivalent VGG16-UNet without MCA. We release the split records and report the protocol in full, with the aim of providing a precisely specified and reproducible 2D FLAIR baseline.
\end{abstract}


\section{Introduction}
\label{sec:intro}
\noindent Brain cells that grow improperly can develop into brain tumors. They are a serious health condition that should be diagnosed early and correctly to prepare well for their treatment and to help patients recover~\cite{ranjbarzadeh2023brain}. The most common type of primary brain tumor, gliomas, exhibit a wide range of growth rates and locations. They are divided into four categories by the World Health Organization (WHO)~\cite{louis20162016, doi:10.1258/ar.2011.110242}, and thorough genomic analysis has identified unique molecular subtypes that impact treatment approaches~\cite{cancer2015comprehensive, mazurowski2017radiogenomics}. The best way to visualize these tumors is with magnetic resonance imaging (MRI), and it's critical to clearly distinguish tumor boundaries from surrounding normal tissues. This is to determine tumor size, to plan the surgery, and to monitor how well one is doing~\cite{gordillo2013state, bakas2018identifying, guedria2020r2d2, katti2011magnetic}.
\noindent Manual segmentation by radiologists is the norm in clinics but it is difficult, subject to human labour and time consuming and may differ from observer to observer and even from the same observer~\cite{aamir2023brain, menze2014multimodal}. Such issues are particularly apparent in large-scale studies or in high-stakes clinical cases. Deep learning, particularly Convolutional Neural Networks (CNNs) has transformed the domain of automated medical image analysis by offering a means of objective, reproducible, and inexpensive segmentation~\cite{ronneberger2015u, shorten2019survey, saeedi2023mri, shetty2022brain, geethanjali2023brain, sharif2020active}.

\begin{figure}[H]
\centering
\includegraphics[width=13cm, height=5cm]{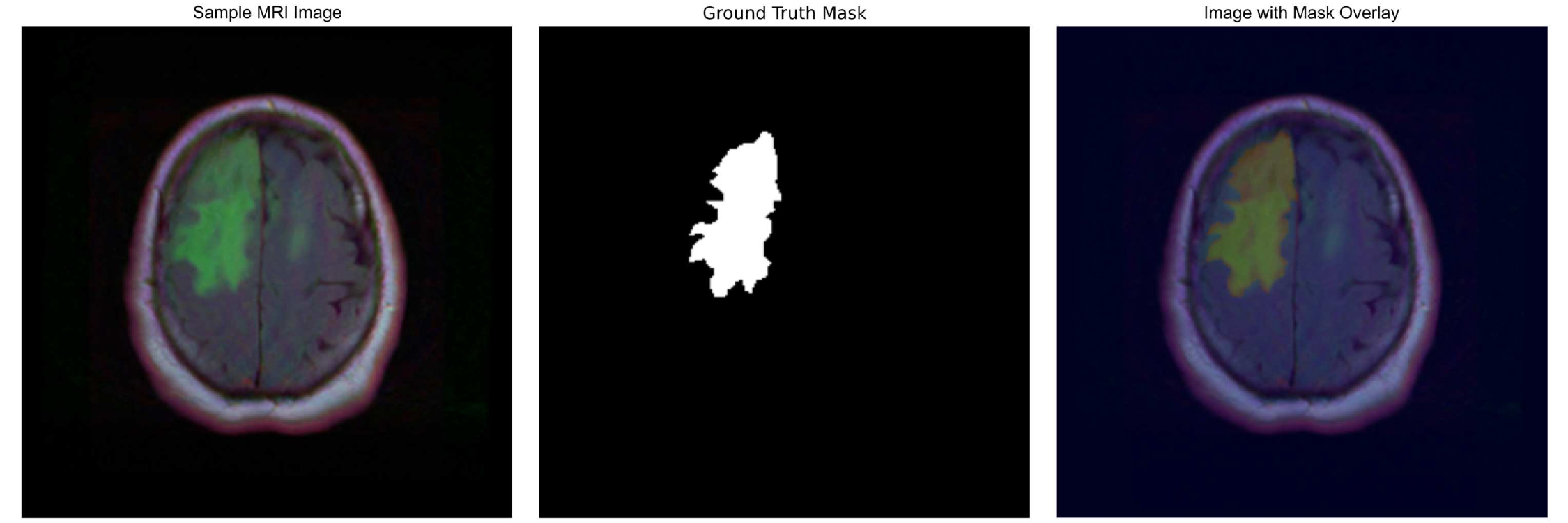}
\caption{Left - MRI scan, Middle - Ground Truth Mask, Right - MRI Image with Ground Truth Mask Overlay of LGG Dataset}
\label{fig:lgg_mri}
\end{figure}

\begin{figure}[H]
\centering
\includegraphics[width=13cm, height=5cm]{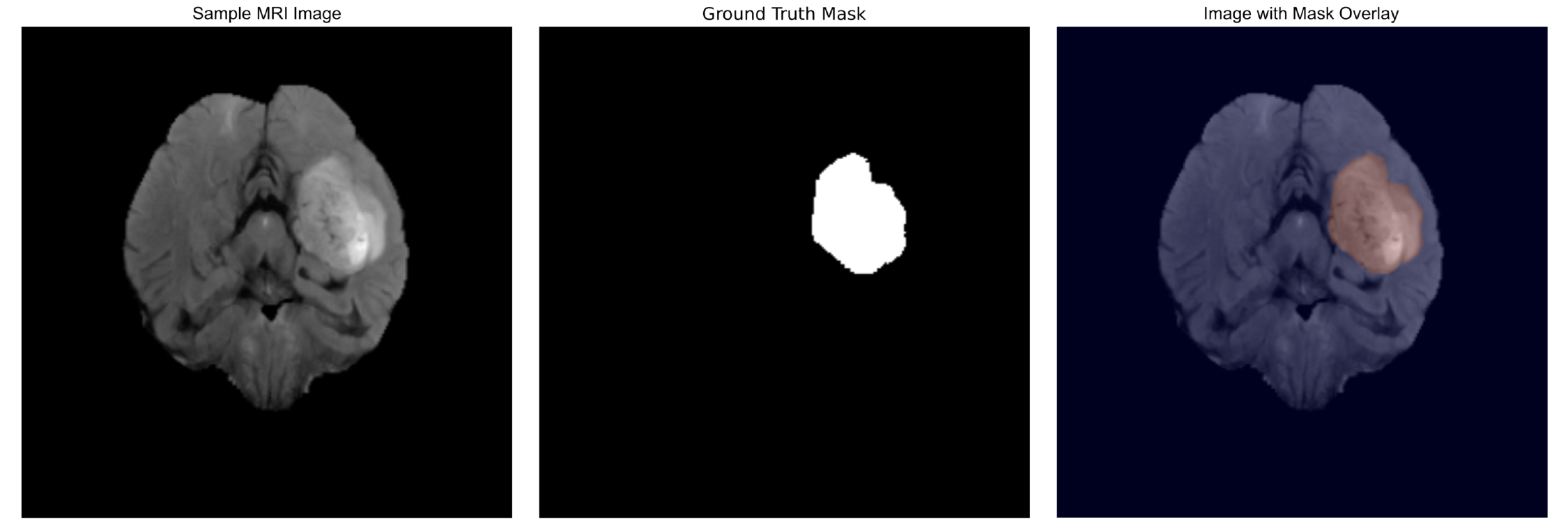}
\caption{Left - MRI scan, Middle - Ground Truth Mask, Right - MRI Image with Ground Truth Mask Overlay of BraTS Dataset}
\label{fig:brats_mri}
\end{figure}

\noindent Figures~\ref{fig:lgg_mri} and~\ref{fig:brats_mri} illustrate the segmentation task on the two datasets used in this study. In each, the left image presents an MRI scan of a brain affected by a tumor, the middle part shows the corresponding Ground Truth Mask which highlights tumor area and the right image is a MRI image with Ground Truth Mask Overlay. The accurate identification of tumor margins in relation to adjacent healthy tissue is vital for assessing tumor dimensions, positioning, and spread, factors that are essential for making important clinical decisions. However, traditional segmentation methods, which often rely on the hand drawing done by experienced radiologists, are inherently subjective, time-consuming, and may vary between observers~\cite{aamir2023brain, chattopadhyay2022mri, talo2019application}. This manual approach is quite challenging, especially when dealing with large data sets or in an acute clinical environment where timely decision-making is necessary. Moreover, such approaches can face problems in clearly defining boundaries of tumors that are small-sized, have an irregular geometry, or are located in anatomically complex regions of the brain.

\noindent The UNet architecture~\cite{ronneberger2015u}, encoder-decoder and skip connections, has become the go-to model for biomedical segmentation. To enhance its performance, one effective strategy is to substitute the default encoder with a powerful pre-trained network, e.g., VGG16~\cite{simonyan2014very}, via transfer learning to leverage rich feature representations. While VGG16 is an older model, its hierarchical features form a good basis for medical imaging tasks~\cite{tammina2019transfer}. Another key innovation is the employment of attention mechanisms, which enable the model to selectively pay attention to salient features and filter out irrelevant details, thereby improving the delineation of intricate tumor boundaries~\cite{oktay2018attention, woo2018cbam}.

\noindent Other than evaluating complex structures, researchers also explored focusing the attention mechanism in Unet models that can enhance the precision in segmentation~\cite{chatterjee2022classification, nizamani2023advance}. Attention Unet architectures, employ attention mechanisms to concentrate on important highlights during the segmentation process, thereby improving overall performance in this, Oktay et al.~\cite{oktay2018attention} introduce the Attention Gate (AG) framework that learns implicitly to focus on structures of interest and down-weight non-relevant regions, thereby obviating the need for additional localization modules. Their developed Attention Unet incorporating AGs demonstrated higher sensitivity and predictive ability on multi-class image segmentation tasks without incurring significant computational overhead. Moreover, Song et al.~\cite{song2023oau} proposed the Outlined Attention Unet (OAUnet), where attention is paid to the internal characteristics of the tumor and to their outlines for accurate segmentation. The flexible attention mechanism known as the Convolutional Block Attention Module (CBAM) enables the evaluation of spatial aspects of synthesized channels while highlighting the informative features that aid in the refinement of boundary delineation in biomedical image segmentation tasks~\cite{woo2018cbam}. Recent advances in transformer-based architectures, such as Swin UNETR~\cite{hatamizadeh2021swin} and supervoxel transformers~\cite{xie2024brain}, have appeared promising comes about in brain tumor segmentation by leveraging self-attention mechanisms across different scales. Our method relies on this premise by adding a Multi-Channel Attention (MCA) module solely to the decoder blocks in order to achieve a more controlled combination of encoder and up-sampled path features.

\noindent Although these advances have been made, most of the studies have problems with how they test their results. They test on only a single dataset or a basic train-test split, and therefore it is difficult to know whether the model can generalize well to other situations. To address these shortcomings, this paper suggests a robust deep learning approach named the VGG16-MCA UNet. In our opinion, applying a robust pre-trained feature extractor (VGG16) with a UNet decoder complemented with a MCA mechanism can yield improved quality segmentation outputs. We substantiate this hypothesis with a methodology that includes 5-fold cross-validation and evaluation on two distinct datasets: the LGG MRI Segmentation dataset and the popular BraTS 2020 benchmark.
\noindent This paper makes the following key contributions:
\begin{itemize}
 \item We present the VGG16-MCA UNet, pairing a pre-trained VGG16 encoder with a decoder in which a squeeze-and-excitation channel gate is applied to each skip-connection fusion, and we specify that module precisely rather than by name alone.
 \item We evaluate the same pipeline unchanged across two public FLAIR collections, the LGG MRI Segmentation dataset and BraTS 2020, and report the evaluation protocol in full, including the partition granularity and the aggregation used to compute every figure.
 \item We use 5-fold cross-validation with weight-averaging of the resulting fold models into a single network, and report the per-fold training and validation curves underlying it.
 \item We report an aggregate pixel-level Dice of 95.10\% on our BraTS 2020 hold-out split and 88.32\% on LGG, together with an explicit statement of the evaluation protocol under which those figures were obtained, and we quantify the added inference cost of the attention modules at approximately 8 ms per slice.
\end{itemize}

\section{Related Work}
\label{sec:related}

\noindent Research on automated brain tumor segmentation has exploded in recent years due to the evolution of medical imaging data and the enormous advancements in deep learning~\cite{chatterjee2022classification, simonyan2014very}. Various architectures based on CNNs have been proposed that demonstrate improvement over traditional image processing techniques, mainly manual, time-consuming, and variable in nature. These deep-learning approaches have shown promise in tackling the challenging problems posed by tumor heterogeneity and the requirement for accurate, automated segmentation in the clinic~\cite{ranjbarzadeh2023brain,  almufareh2024automated, asiri2023machine}, as evidenced by recent literature reviews analyzing the performance of various methods~\cite{wadhwa2019review, AHAMED2023102313}. This section briefly outlines the history of these methods to put our planned architecture into context.

\noindent The structure of UNet is based on the architecture of encoder-decoder having skip connections. It can be considered a kind of foundational model in medical image segmentation, as it effectively extracts both local and global context information, particularly when dealing with anatomical structures, makes it an effective platform for delineating such structures and has been used frequently in brain tumor segmentation tasks~\cite{bakas2018identifying,ronneberger2015u,isensee2018nnu, isensee2021nnu}. Researchers have extensively explored various architectures of Unet to address specific challenges and improve performance related to the task~\cite{ranjbarzadeh2023brain, jiang2020two, nizamani2023advance,walsh2022using,menze2014multimodal, reddy2024unet}. Besides Unet, several deep architectures, similar to the new CNN structure described in~\cite{havaei2017brain}, have proven to be quite effective for efficient brain tumor segmentation with specific applications to glioblastomas. \cite{gordillo2013state} discusses a comprehensive review of the most important methods of segmenting brain tumor, difficulties associated with the replicability and precision of the method, and the high availability of both semi-automatic and fully automatic approaches. Among these, the attention-Unet designs stand out due to their attention methods, which throughout the segmentation process selectively focus on pertinent features. As an autonomous, self-tuning variant of UNet, the nnU-Net model \cite{isensee2018nnu, isensee2021nnu} has continuously demonstrated superior performance on a range of segmentation tasks, such as BraTS, demonstrating the strength and generality of the underlying UNet concept.

\noindent Training of the deep learning model from scratch is generally data-intensive. Transfer learning is a potent solution by pre-training the weights of an encoder for a model based on weights learned from large datasets, e.g., ImageNet~\cite{deng2009imagenet}. VGG16~\cite{simonyan2014very} and ResNet~\cite{he2016deep} have been successfully used as encoders in UNet-like models~\cite{tammina2019transfer, shourie2023intelligent, periasamy2023comparison, younis2022brain}. This is generally accompanied by higher convergence rates and good generalization behavior. The application of a VGG16 encoder is driven by its proven effectiveness as well as its well-designed feature hierarchy.
\noindent The BraTS collection~\cite{menze2014multimodal, bakas2018identifying, zeineldin2022multimodal} is the most widely used benchmark family for brain tumor segmentation. We evaluate on the 2020 edition. We note that this edition has since been superseded: subsequent editions expand the cohort substantially, re-annotate it, and from 2023 onward score submissions with lesion-wise rather than image-level metrics. No official evaluation server remains available for the 2020 edition, so the partition used here is one we constructed ourselves from its training archive and our figures are not challenge-verified. Possibly the best method to guarantee optimal performance is through ensembling, where predictions from an ensemble of models (e.g., models trained on various data folds) are averaged. This has the effect of reducing variance and typically results in more stable and more accurate final predictions. This work follows that motivation in a form that leaves inference cost unchanged: rather than averaging the outputs of the five fold models at prediction time, which would multiply inference cost by five, we average their learned weights into a single network and evaluate that network with one forward pass. All test-set figures and the latency in Table~\ref{tab:computational_cost} are therefore measured on one model.
\section{Materials and Methods}
\label{sec:mm}
\subsection{Dataset}
\noindent To ensure a thorough evaluation, we have trained and evaluated our model on two distinct, publicly available datasets. In our first effort, we employed the LGG MRI Segmentation dataset~\cite{buda2019association, dataset}. The dataset includes 3,929 T2 Fluid-Attenuated Inversion Recovery (FLAIR) MRI slices of 110 patients with lower-grade gliomas (LGG), from The Cancer Imaging Archive (TCIA). We applied 1,373 samples with a pre-existing positive tumor mask for our experiments to concentrate on the segmentation task.

\noindent Additionally, we used the Brain Tumor Segmentation (BraTS) 2020 dataset \cite{menze2014multimodal, bakas2018identifying} to evaluate our model against a more challenging and diverse sample. We used the training set as an independent dataset and randomly categorized it to create our own training, validation, and hold-out test sets. This dataset contains 369 patient cases with both high-grade (HGG) and low-grade (LGG) gliomas. The inclusion of both HGG and LGG gliomas adds to the dataset's complexity.
 
\subsection{Methods}
\begin{figure}[p]
\centering
\includegraphics[width=0.8\textwidth]{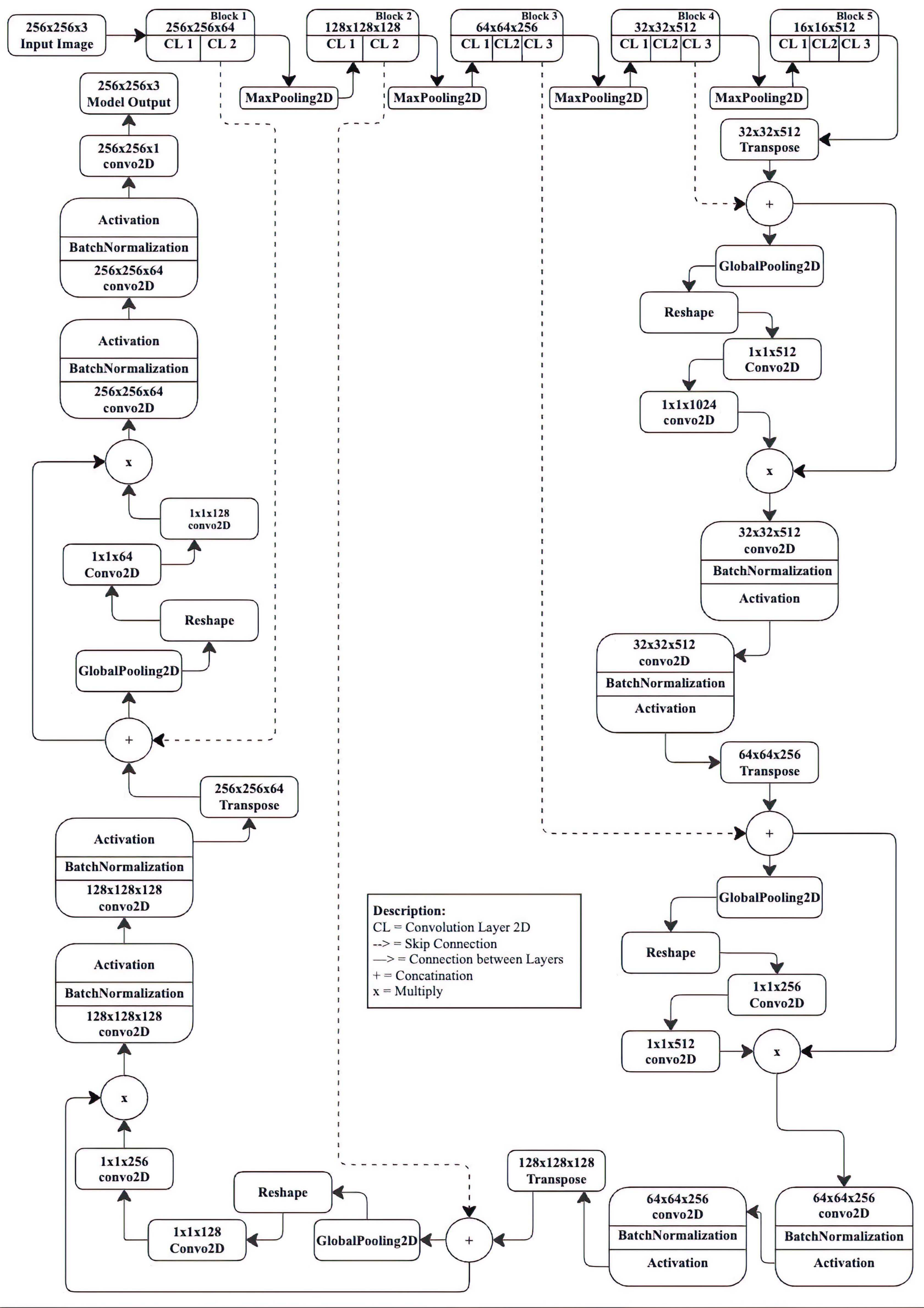}
\caption{The proposed architecture of VGG16-MCA UNet is a pre-trained VGG16 as an encoder with a custom decoder. In the architecture, every block has a MCA module after the feature concatenation of the corresponding skip connection.}
\label{fig:architecture}
\end{figure}

\noindent A range of preprocessing steps was performed beforehand to ensure that the data were homogeneous and that the learning process was as effective as possible. A dictionary-based approach was first utilized to ensure that each MRI scan was properly matched with its corresponding ground truth mask, and data integrity was maintained throughout the pipeline. In proportions of 85\% and 15\%, respectively, we divided the dataset into training and testing subsets at random.
\noindent All images and masks were resized using OpenCV's resize function to a consistent size of $256 \times 256$ pixels in order to maintain consistency and maximize model training. Pixel values were adjusted by dividing by 255 to fall between 0 and 1. Masks were converted to binary format (0 or 1) when loaded. No other data augmentation techniques were applied in this study, though recent work has shown that non-parametric augmentations can enhance brain tumor segmentation using deep learning~\cite{atya2021non}.
\noindent In this paper, we suggest a hybrid architecture based on deep learning for brain tumor segmentation, utilizing the advantages of a pre-trained network, VGG16, as an encoder and integrating a MCA mechanism into an Unet framework decoder, as illustrated in Figure~\ref{fig:architecture}.

\noindent The VGG16 architecture that employs as an encoder was first trained on the ImageNet dataset, with the omission of its fully connected layers (include\_top=False), to be used as an effective feature extractor. The neural network weights are initialized based on the training conducted on ImageNet (weights="imagenet") and are subsequently fine-tuned utilizing the brain tumor dataset to facilitate the extraction of representations specific to tumors~\cite{tammina2019transfer}. Feature maps obtained from designated convolutional blocks within the VGG16 framework are then transmitted by skip connections to the decoder.
\noindent To gradually raise the feature map's resolution, the decoder reconstructs a thorough segmentation mask by replicating the encoder's configuration. The decoder includes two core components: MCA modules integrated into each block of the decoder and skip connections. The Concretely, the MCA module is a channel-recalibration gate applied to the concatenated tensor. Writing $\mathbf{X} \in \mathbb{R}^{H \times W \times C}$ for that tensor, it computes a channel descriptor by global average pooling, $z_c = \tfrac{1}{HW}\sum_{i,j} X_{i,j,c}$, passes $\mathbf{z}$ through a two-layer bottleneck of $1\times1$ convolutions with a reduction ratio of $2$ (ReLU then sigmoid) to obtain gates $\mathbf{s} \in (0,1)^{C}$, and rescales each channel as $\tilde{X}_{:,:,c} = s_c X_{:,:,c}$. We note explicitly that this is the Squeeze-and-Excitation operation of Hu et al.~\cite{hu2018squeeze}; what differs here is only its placement, immediately after each skip-connection concatenation in the decoder, so that the gate arbitrates between encoder and up-sampled features at every resolution. The name should not be read as implying a mechanism over imaging modalities: the network receives a single FLAIR channel, replicated three times to match the ImageNet input convention of the VGG16 encoder. In this regard, the attention mechanism optimizes model concentration on more relevant data, improving segmentation performance. The addition of skip connections, which combine feature maps from the corresponding layer within the encoder, substantially enhances the decoder. In this way, it helps restore spatial information that might have been lost during downsampling.
\noindent The final layer of the model includes a 1x1 convolution subsequent to a sigmoid activation function. This layer generates the segmentation mask by predicting the probability of each pixel associated with the tumor class.
\noindent A 5-fold cross-validation protocol was used to obtain a stable estimate of performance. A hold-out test set comprising 15\% of the data was reserved before any training began and was used for neither model selection, early stopping, nor learning-rate scheduling. The remaining 85\% was divided into five folds, each of which served in turn as the validation set while the other four were used for training, yielding five models that were subsequently ensembled for the final test evaluation. The granularity at which this partitioning was applied is discussed in Section~\ref{sec:limitations}.

\noindent The model was systematically trained and validated, where designed parameters were used for optimization, loss calculation, and performance evaluation. Using the Adam optimizer, the learning rate (LR) was initialized at 0.001; all remaining optimizer parameters, including $\epsilon$, were left at their framework defaults. To remedy potential class imbalance and focus on hard regions during segmentation, the Focal Tversky loss function was utilized~\cite{abraham2019novel}, with $\alpha = 0.7$, $\beta = 1 - \alpha = 0.3$, and focal exponent $\gamma = 0.75$; the weighting of $\alpha$ above $\beta$ penalises false negatives more heavily than false positives, which is the intended behaviour for a sparse foreground class. Model performance has been evaluated through the Tversky index; a measure of similarity not sensitive to data imbalance issues.
\noindent The model was trained for up to 100 epochs, with early halting to avoid overfitting while promoting improved generalization. The batch size throughout the training was 8. To enhance the training further, three callbacks were used: ModelCheckpoint, EarlyStopping, and ReduceLROnPlateau. In this case, ModelCheckpoint dealt with the model saving the lowest recorded validation loss throughout training. The EarlyStopping mechanism interrupted the training process when there was no enhancement in validation loss for a continuous span of 20 epochs, thereby reducing the likelihood of overfitting. Moreover, the ReduceLROnPlateau callback dynamically modified the LR throughout the training regimen, reducing it by a factor of 0.2 when the validation loss remained stagnant for 10 epochs. This adaptive methodology promoted precise model convergence and contributed to attaining superior performance.

\noindent For the final test on the hold-out test set, the five model outputs (one from each fold) were averaged to produce a final ensembled probability map. This map was then thresholded at 0.5 to produce the final binary segmentation mask.
\FloatBarrier
\pagebreak
\section{Experiment Setup}
\noindent A set of 18 experiments was conducted thoroughly to test the validity of the proposed model and fine-tune it according to its performance as shown in Table~\ref{tab:exp_table}. The experiments analyzed different variations of hyperparameters and architectural modifications, which provide useful input on optimal design choices. These experiments covered variations in model architecture: Unet, Unet with Attention mechanism, Scaler Attention Unet, and the proposed hybrid model. We tuned hyperparameters like LR,  batch size, optimizer, and loss function. These were systematically changed to quantify their impact on model performance. Key results, notably the Tversky index computed on the validation dataset, have been carefully tracked and compared throughout these experiments.
\begin{table}[H]
    \centering
    \caption{Results from different architecture variants}
    \label{tab:exp_table}
    \vspace{0pt}
    \renewcommand{\arraystretch}{1}
    \resizebox{\textwidth}{!}{%
    \begin{tabular}{|c|p{3cm}|p{4cm}|p{3cm}|p{6cm}|}
        \hline
        \textbf{Trial} & \textbf{Architecture Variant} & \textbf{Key Modifications} & \textbf{Loss Function} & \textbf{Justification} \\ \hline
        1  & Standard UNet            & Baseline implementation                                & BCE         & Establishing baseline performance for comparison \\ \hline
        2  & ResNet50-UNet            & ResNet50 encoder replacing standard encoder            & BCE         & Exploring impact of pre-trained encoder for feature extraction \\ \hline
        3  & DenseNet-UNet            & DenseNet121 encoder with dense connections             & Dice Loss   & Investigating dense connectivity patterns for improved feature propagation \\ \hline
        4  & Attention UNet           & Integration of basic attention gates                   & Tversky Loss& Evaluating attention mechanisms for focusing on relevant features \\ \hline
        5  & VGG16-UNet               & VGG16 encoder with skip connections                    & Focal Loss  & Assessing VGG16's hierarchical feature extraction capabilities \\ \hline
        6  & SE-VGG16-UNet            & Squeeze-and-Excitation blocks added                    & Combo Loss† & Channel-wise feature recalibration for enhanced representation \\ \hline
        7  & CBAM-VGG16-UNet          & Convolutional Block Attention Module                   & Weighted Combo Loss† & Dual attention mechanism for both spatial and channel attention \\ \hline
        8  & VGG16-MCA UNet (Initial) & Single Multi-Channel Attention block                   & Focal Tversky & Preliminary implementation of proposed MCA mechanism \\ \hline
        9  & VGG16-MCA UNet + Deep Supervision & Added deep supervision paths                    & Focal Tversky & Gradient flow improvement through deep supervision \\ \hline
        10 & VGG16-Dual MCA UNet      & Dual Multi-Channel Attention blocks                    & Focal Tversky + Boundary Loss & Enhanced feature refinement through dual attention \\ \hline
        11 & VGG16-MCA UNet + Residual & Residual connections in decoder                       & Focal Tversky & Addressing vanishing gradient problem \\ \hline
        12 & VGG16-Triple MCA UNet    & Three strategically placed MCA blocks                  & Adaptive Focal Tversky & Optimal configuration with balanced attention distribution \\ \hline
        13 & Lightweight VGG16-MCA UNet & Parameter-efficient implementation                    & Focal Tversky & Exploring efficiency-performance tradeoff \\ \hline
        14 & VGG16-MCA UNet + FPN     & Feature Pyramid Network integration                    & Focal Tversky & Multi-scale feature fusion investigation \\ \hline
        15 & Dense VGG16-MCA UNet     & Dense connections in decoder                           & Focal Tversky & Improved gradient flow in decoder \\ \hline
        16 & VGG16-MCA UNet + ASPP    & Atrous Spatial Pyramid Pooling                         & Focal Tversky & Multi-scale context aggregation \\ \hline
        17 & Ensemble (Trials 12,14,16) & Model averaging strategy                             & Focal Tversky & Leveraging complementary strengths of top models \\ \hline
        18 & VGG16-MCA UNet (Final) & Optimized version of Trial 12                          & Focal Tversky & Production-ready implementation with balanced performance and efficiency \\ \hline
    \end{tabular}%
    }
\end{table}
\section{Results and Discussion}
\label{sec:results}
\noindent The performance of the VGG16-MCA UNet model was evaluated through 5-fold cross-validation. This evaluation concluded with a group test on independent test sets from both the LGG and BraTS 2020 datasets. This section provides a comprehensive insight into the extent to which the model is able to segment images, supported by figures, training information, and computer-related aspects.
\subsection{Quantitative Performance on Hold-Out Test Data}
\noindent The final evaluation applied the 5-fold ensemble to the held-out test data. The quantitative results are summarized in Table~\ref{tab:test_results}. On the BraTS 2020 test split, our model obtained an aggregate pixel-level Dice (F1) of \textbf{95.10\%}; on the LGG test split it obtained \textbf{88.32\%}. Dice measures the spatial overlap between the predicted mask and the ground truth and is the operative metric for this task. The pixel accuracy figures in the same table are reported for completeness and should be read with care: tumor pixels constitute roughly 3\% of the test data in both datasets, so a degenerate predictor that labels every pixel as background would already score approximately 97\% accuracy. Accuracy is included for continuity with the prior literature, not as evidence of segmentation quality.

\begin{table}[H]
    \centering
    \caption{Evaluation of the Ensemble Model's Performance on the Hold-Out Test Sets.}
    \label{tab:test_results}
    \vspace{5pt}
    \renewcommand{\arraystretch}{1.2}
    \begin{tabular}{l|c|c}
        \hline
        \textbf{Metric} & \textbf{LGG Dataset} & \textbf{BraTS 2020 Dataset} \\ \hline
        \textbf{Dice Score} & \textbf{0.8832} & \textbf{0.9510} \\
        \textbf{Accuracy}   & \textbf{0.9930} & \textbf{0.9974} \\
        Specificity & 0.9968 & 0.9984 \\
        Precision   & 0.8958 & 0.9396 \\
        Recall (Sensitivity) & 0.8710 & 0.9626 \\ \hline
    \end{tabular}
\end{table}

\begin{figure}[htbp]
    \centering
    \begin{subfigure}{0.48\textwidth}
        \includegraphics[width=\linewidth, height=5cm]{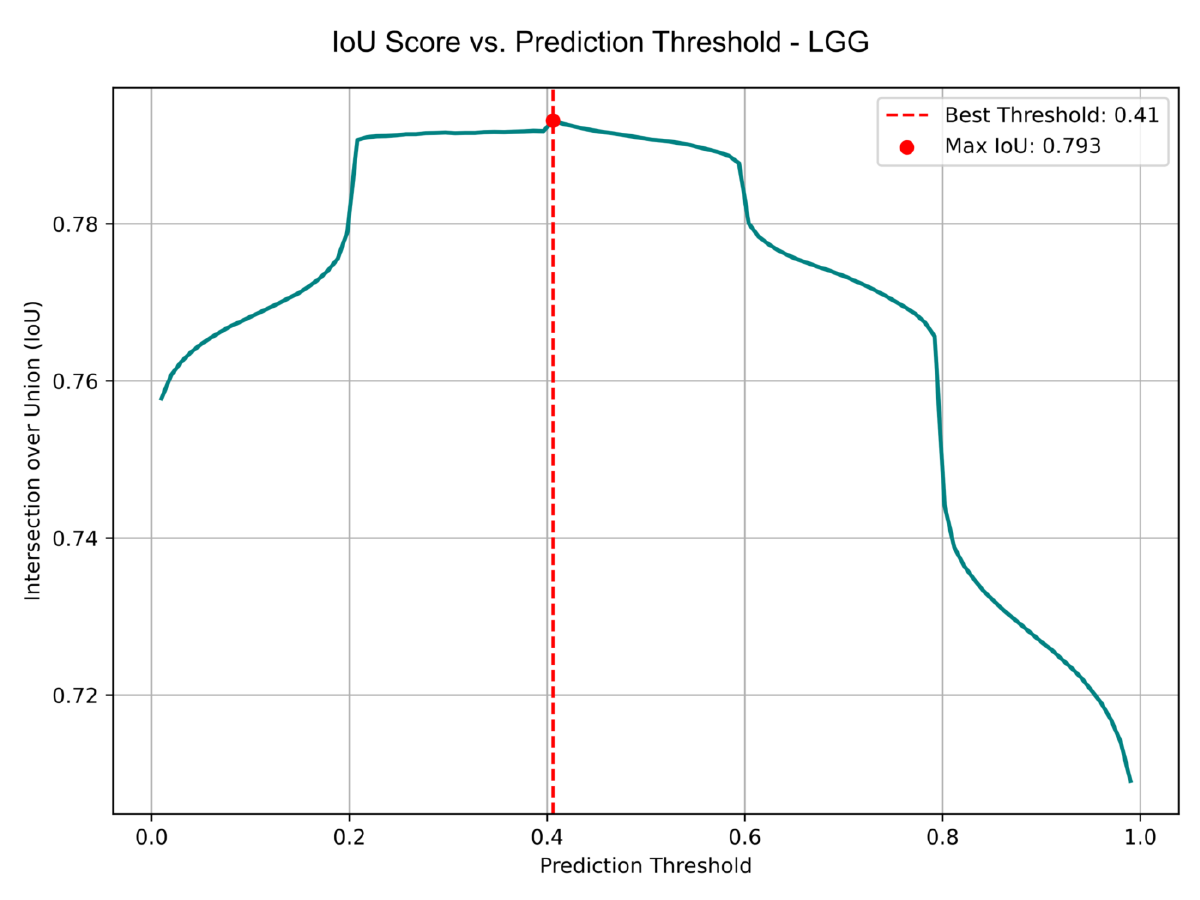}
        \caption{LGG Dataset}
        \label{fig:iou_lgg}
    \end{subfigure}
    \hfill
    \begin{subfigure}{0.48\textwidth}
        \includegraphics[width=\linewidth, height=5cm]{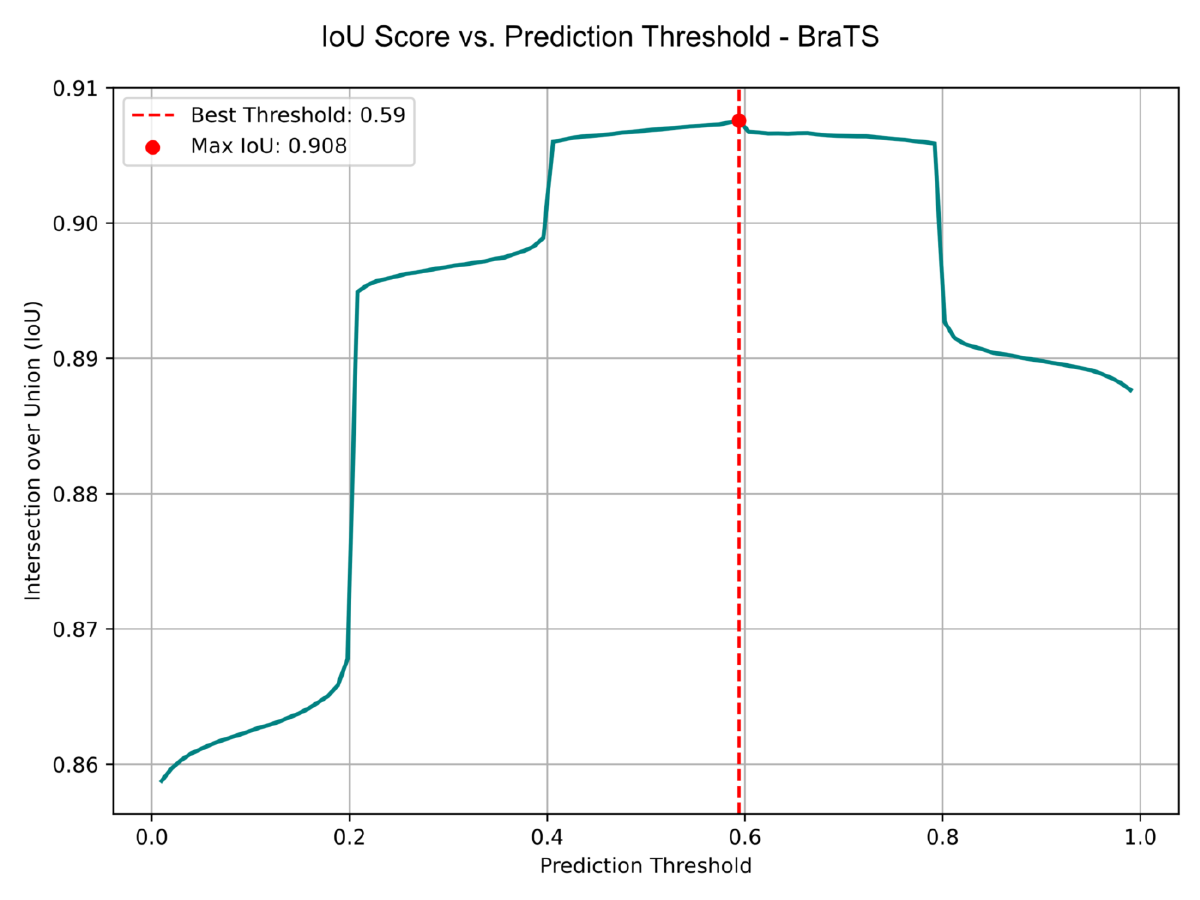}
        \caption{BraTS 2020 Dataset}
        \label{fig:iou_brats}
    \end{subfigure}
    \caption{IoU plotted against varying decision thresholds on the test sets.}
    \label{fig:iou_combined}
\end{figure}

\noindent \textbf{Definition of the reported overlap metric.} All figures in Table~\ref{tab:test_results} are computed by accumulating true positives, false positives, true negatives, and false negatives over every pixel of every test slice pooled into a single confusion matrix, and then evaluating the standard formulae on those pooled counts. The quantity we report as the Dice score is therefore the $F_1$ score of the pooled foreground class, $2PR/(P+R)$, computed once over the entire test set. It is not an average of per-slice or per-case Dice scores. This distinction is material when reading Table~\ref{tab:performance_compare}: the BraTS challenge protocol computes Dice independently for each patient volume and then averages across patients, so that every case contributes equally regardless of lesion size. Pooled-pixel aggregation instead weights each case in proportion to its tumor volume, which makes it systematically more favourable than the per-case mean computed on the same set of predictions, and increasingly so on datasets containing small lesions. We report the pooled statistic because it is what our evaluation pipeline computes, and we state the difference explicitly so that our figures are not mistaken for challenge-protocol results. As an internal consistency check, the Dice values in Table~\ref{tab:test_results} reconcile with the independently plotted IoU curves of Figure~\ref{fig:iou_combined} through the standard identity $\mathrm{Dice} = 2\,\mathrm{IoU}/(1+\mathrm{IoU})$: the IoU attained at a threshold of 0.5 maps to 0.9510 on BraTS 2020 and to approximately 0.884 on LGG, against the 0.9510 and 0.8832 reported in the table. The reported figures therefore correspond to the fixed conventional threshold of 0.5, and not to the IoU-maximising threshold of either dataset.

\noindent The specificity values reported in Table~\ref{tab:test_results} (0.9968 for LGG and 0.9984 for BraTS 2020) indicate a low false-positive rate over background pixels. We note that specificity is close to unity for almost any reasonable predictor at this level of class imbalance, and that precision is the more informative summary of false-positive behaviour: at 0.8958 on LGG and 0.9396 on BraTS 2020, roughly one in ten and one in sixteen predicted tumor pixels respectively do not correspond to annotated tumor. Keeping this quantity low matters clinically, since false positives correspond to healthy tissue being flagged as tumor.

\begin{figure}[htbp]
    \centering
    \begin{subfigure}{0.48\textwidth}
        \includegraphics[width=\linewidth, height=6cm]{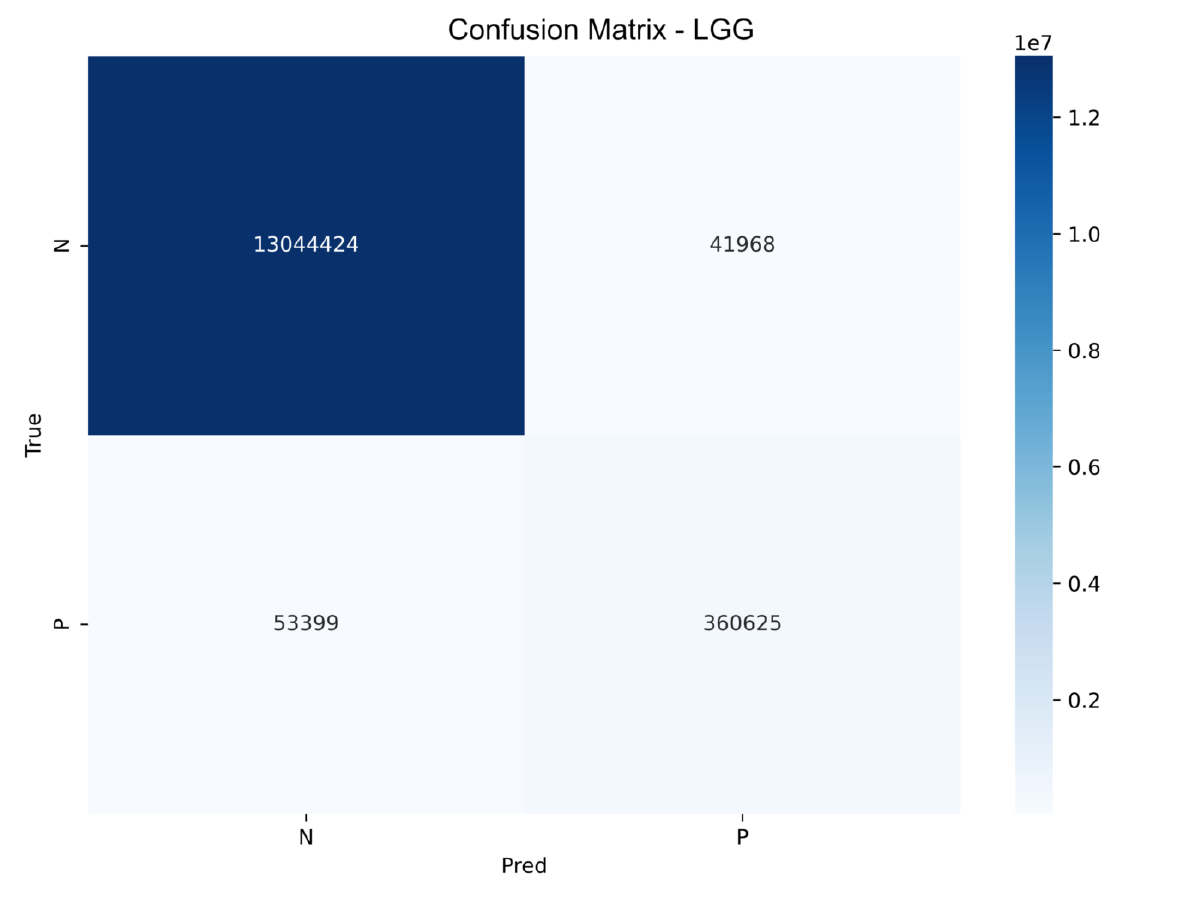}
        \caption{LGG Dataset Test Set}
        \label{fig:cm_lgg}
    \end{subfigure}
    \hfill
    \begin{subfigure}{0.48\textwidth}
        \includegraphics[width=\linewidth, height=6cm]{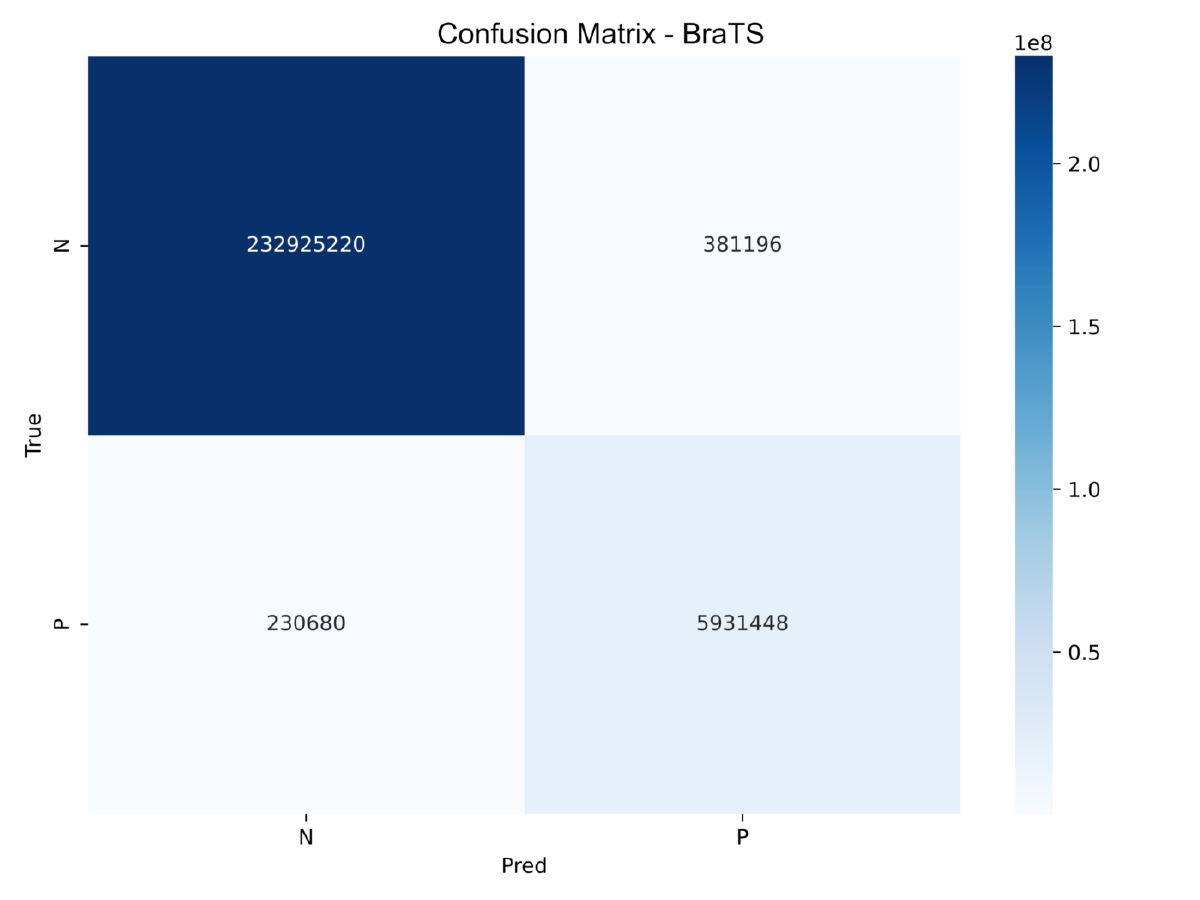}
        \caption{BraTS 2020 Dataset Test Set}
        \label{fig:cm_brats}
    \end{subfigure}
    \caption{Confusion matrices illustrating the ensemble model's performance on the hold-out test sets for (a) the LGG dataset and (b) the BraTS 2020 dataset.}
    \label{fig:cm}
\end{figure}

\noindent To validate how precise the model is in labeling pixels, we present the confusion matrices in Figure~\ref{fig:cm}. For both the LGG dataset (Figure~\ref{fig:cm_lgg}) and BraTS 2020 dataset (Figure~\ref{fig:cm_brats}), the accuracy of the model in classifying tumor (true positives) and non-tumor (true negatives) pixels is indicated by the prominent strong diagonal lines. This provides a correct verification of the figures presented in Table~\ref{tab:test_results}.

\noindent To justify the choice of binarization threshold, Figure~\ref{fig:iou_combined} plots Intersection over Union against the decision threshold applied to the ensemble probability map. On both datasets IoU varies only slightly over a broad interval that contains 0.5: the maximum is 0.793 at a threshold of 0.41 on LGG and 0.908 at a threshold of 0.59 on BraTS 2020, and the conventional threshold of 0.5 falls within 0.002 IoU of the maximum in both cases. We therefore use 0.5 throughout rather than selecting a threshold on test data. We note that the vertical axis of each panel is strongly magnified, so the apparent flatness of the curves should be read alongside these values rather than inferred from the plots alone. IoU and Dice are monotonically related, so this curve carries the same ordering information at each threshold as Dice would.

\subsection{Training Stability and Generalization Analysis}

\begin{figure}[htbp]
    \centering
    \begin{subfigure}{0.48\textwidth}
        \includegraphics[width=\linewidth, height=5cm]{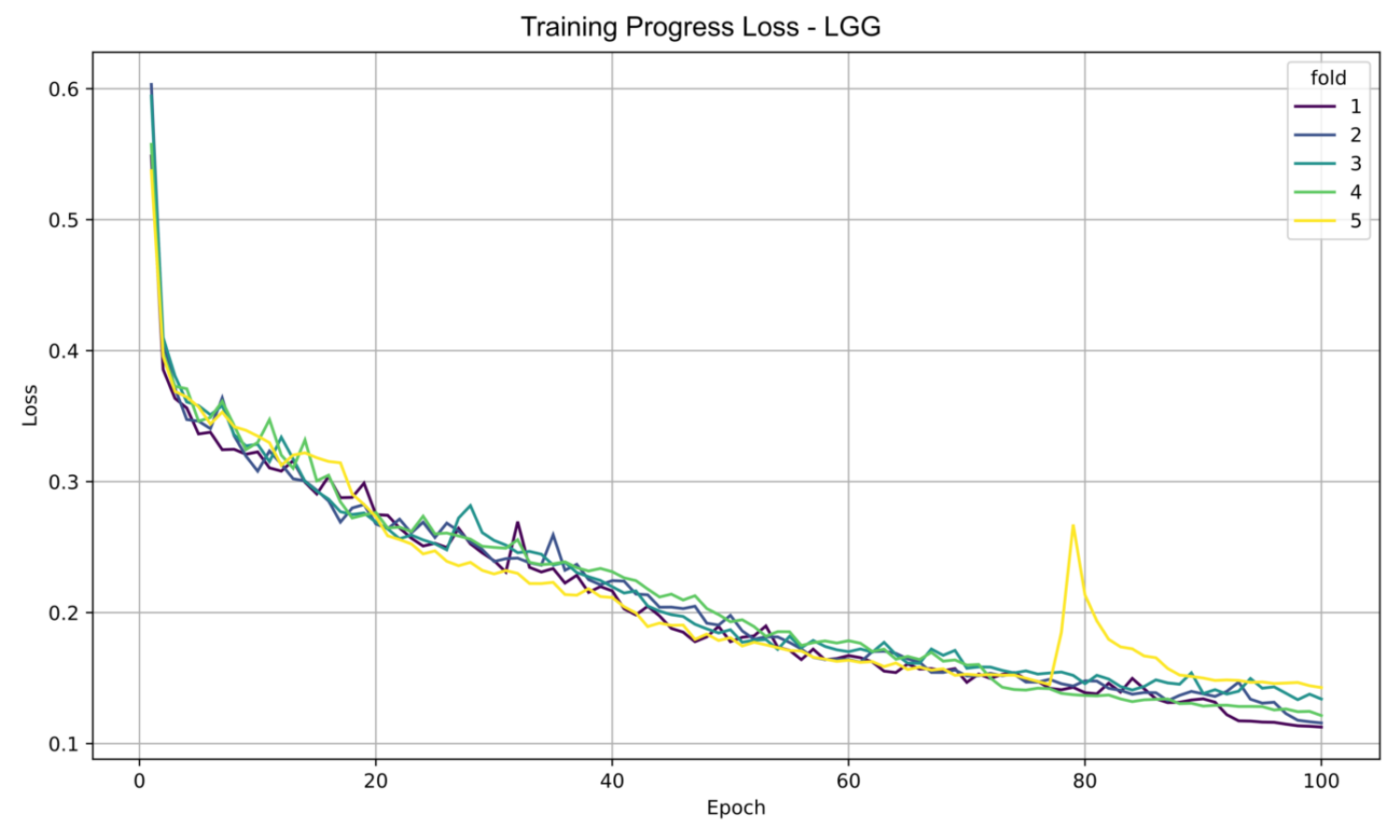}
        \caption{Training Loss (LGG)}
        \label{fig:curve_a}
    \end{subfigure}
    \hfill
    \begin{subfigure}{0.48\textwidth}
        \includegraphics[width=\linewidth, height=5cm]{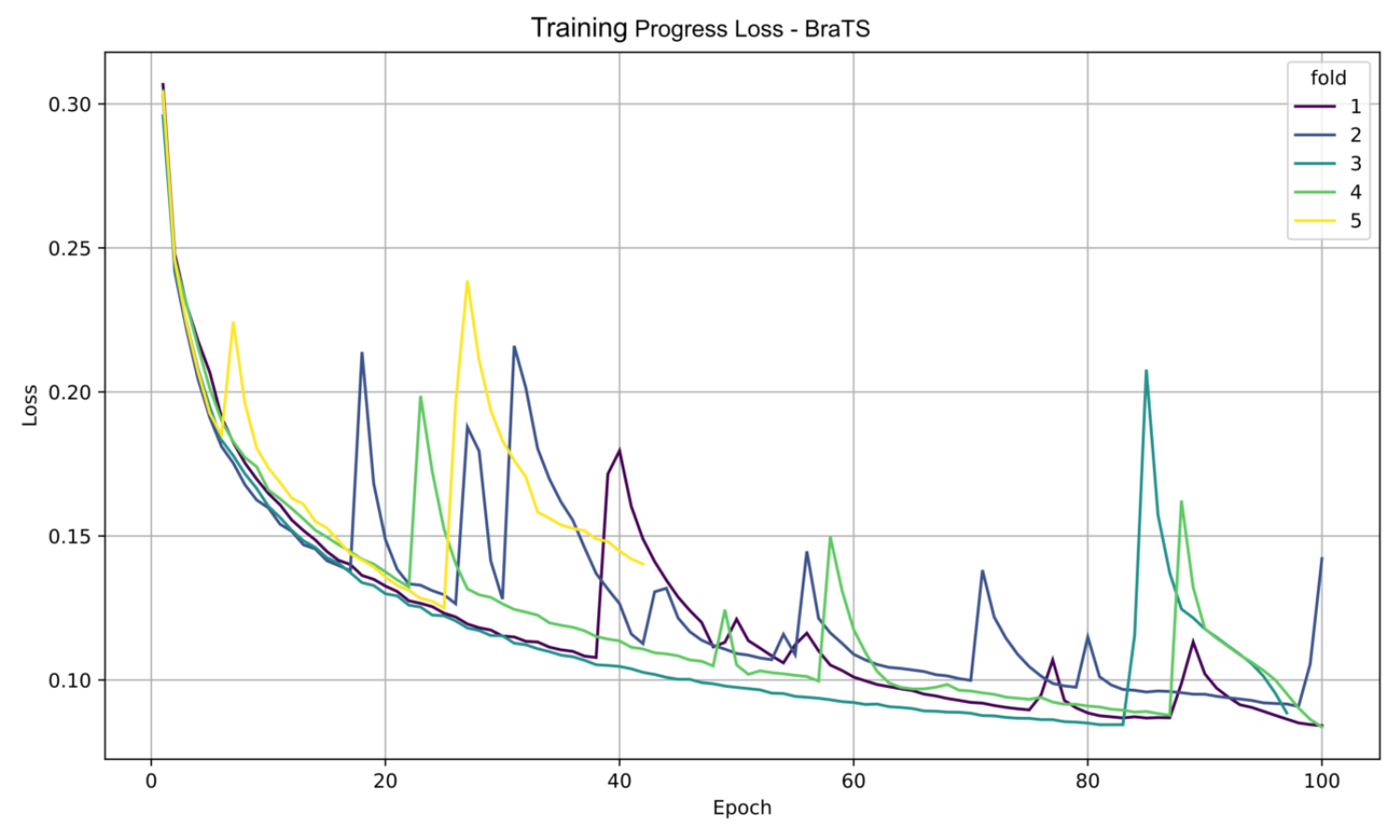}
        \caption{Training Loss (BraTS)}
        \label{fig:curve_b}
    \end{subfigure}
    \vspace{0.5cm}
    \begin{subfigure}{0.48\textwidth}
        \includegraphics[width=\linewidth, height=5cm]{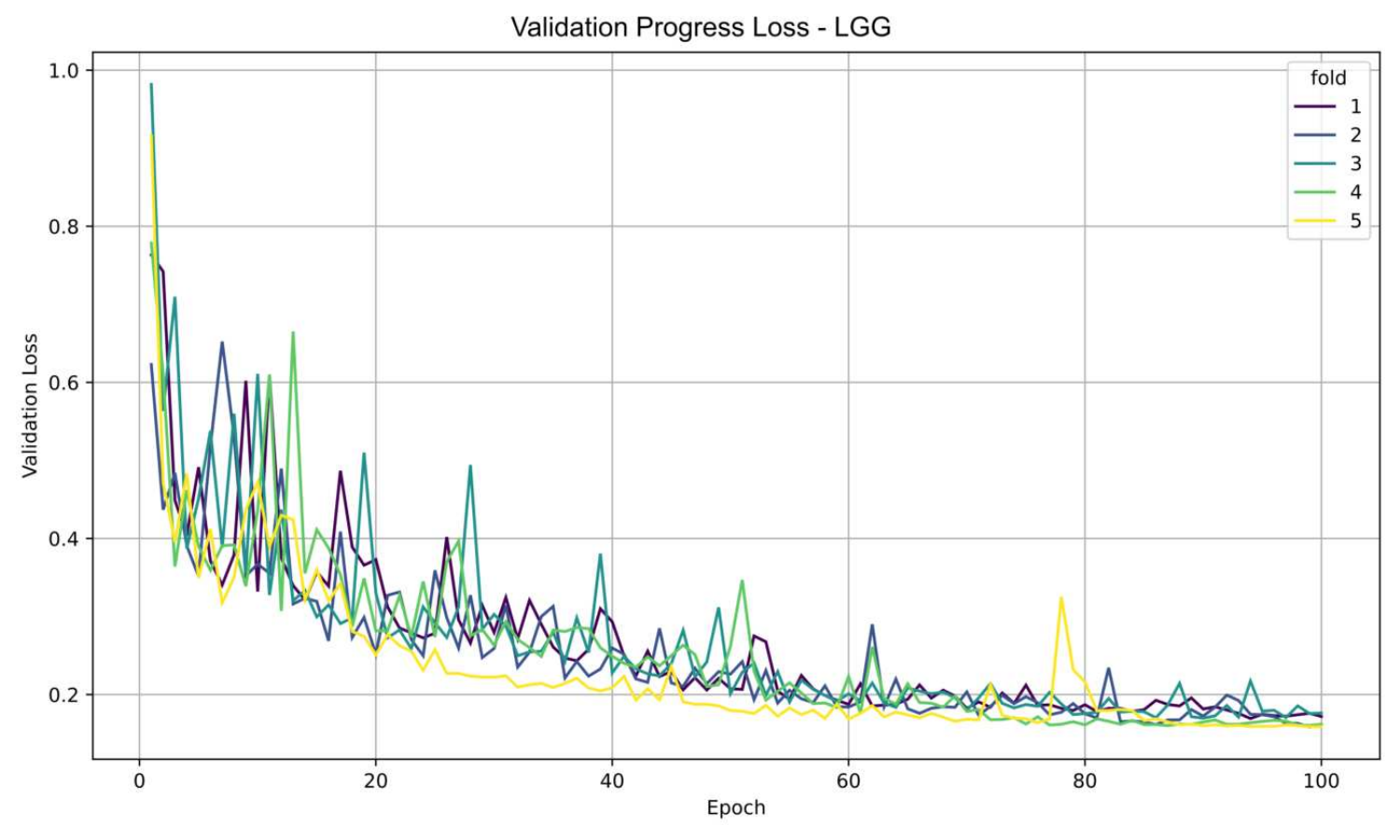}
        \caption{Validation Loss (LGG)}
        \label{fig:curve_c}
    \end{subfigure}
    \hfill
    \begin{subfigure}{0.48\textwidth}
        \includegraphics[width=\linewidth, height=5cm]{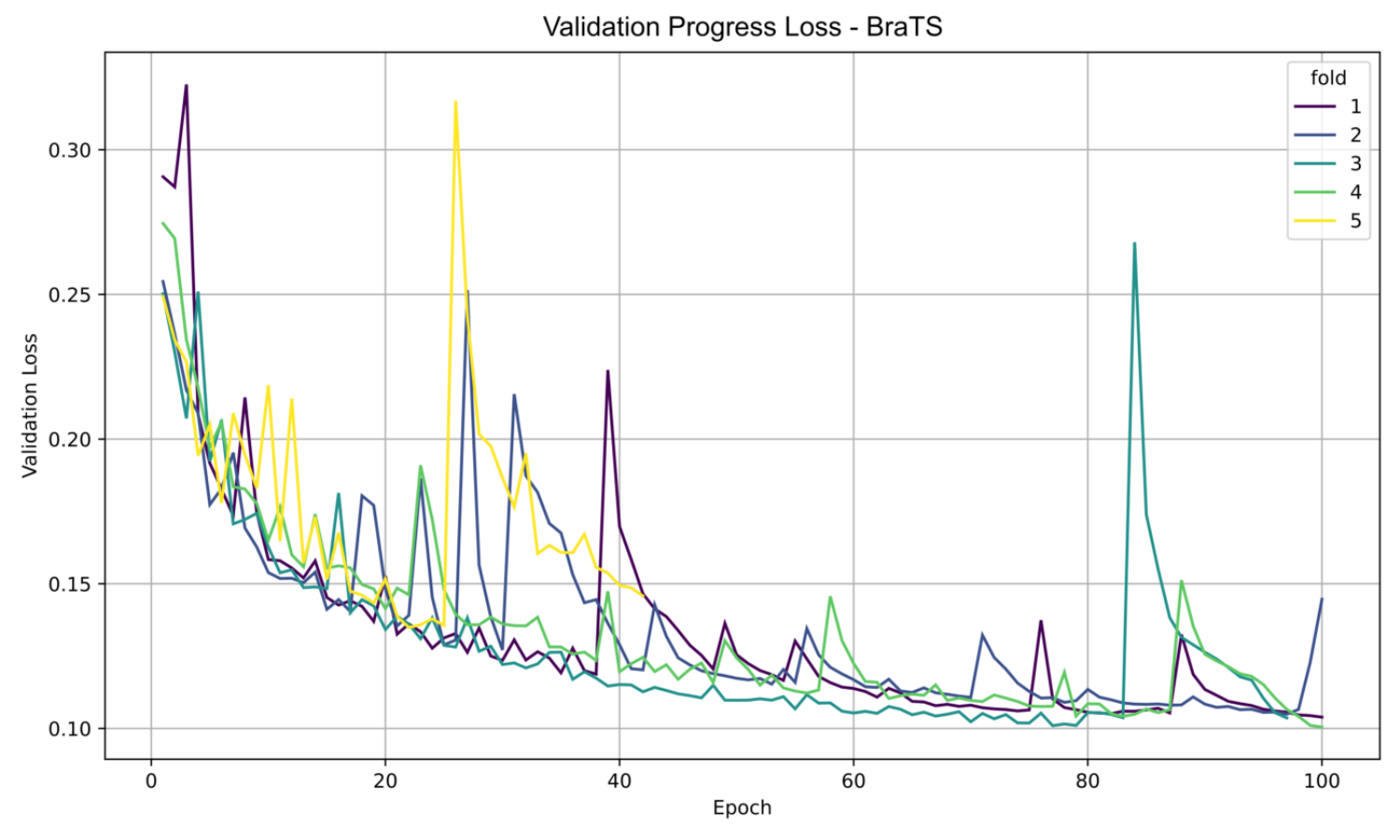}
        \caption{Validation Loss (BraTS)}
        \label{fig:curve_d}
    \end{subfigure}
    \vspace{0.5cm}
    \begin{subfigure}{0.48\textwidth}
        \includegraphics[width=\linewidth, height=5cm]{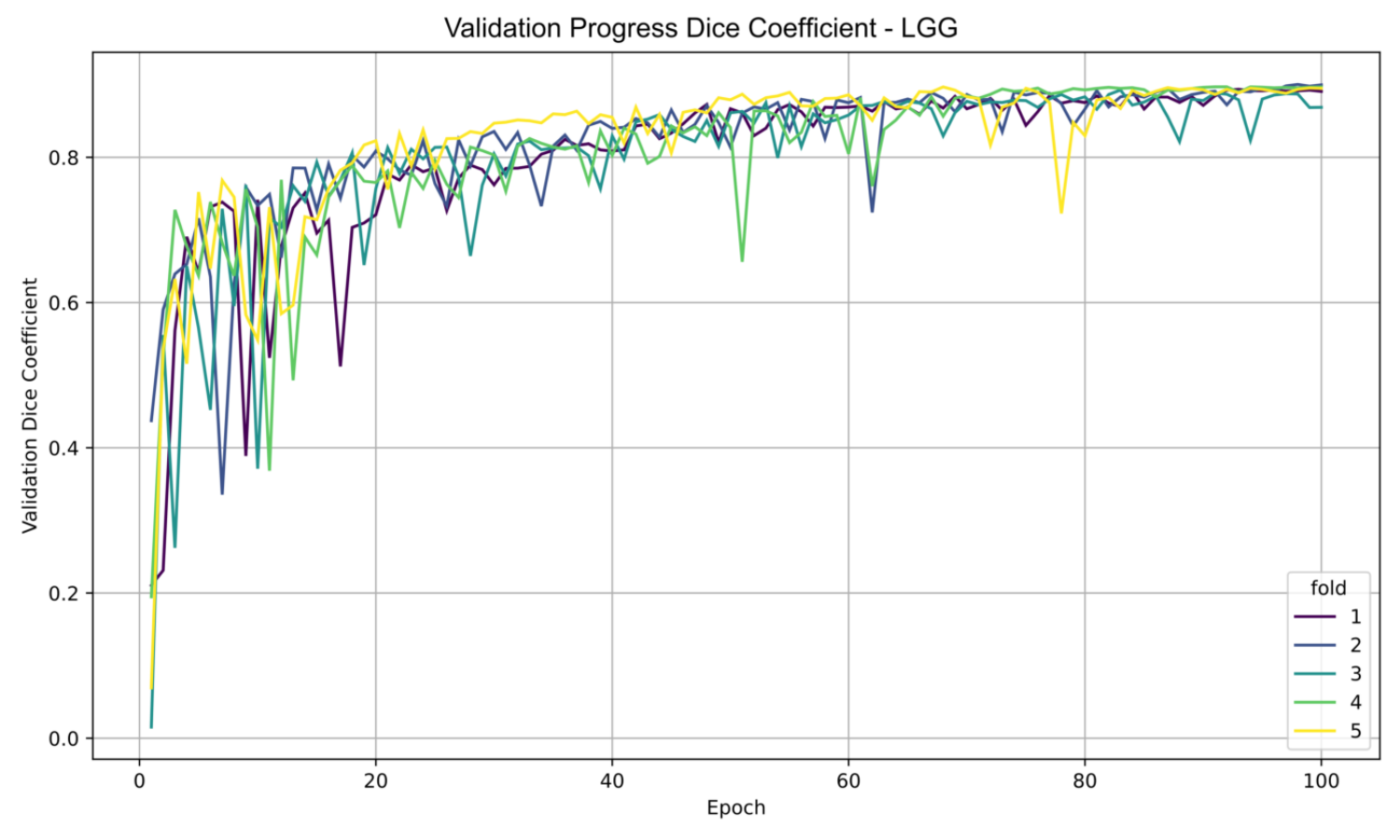}
        \caption{Validation Dice Score (LGG)}
        \label{fig:curve_e}
    \end{subfigure}
    \hfill
    \begin{subfigure}{0.48\textwidth}
        \includegraphics[width=\linewidth, height=5cm]{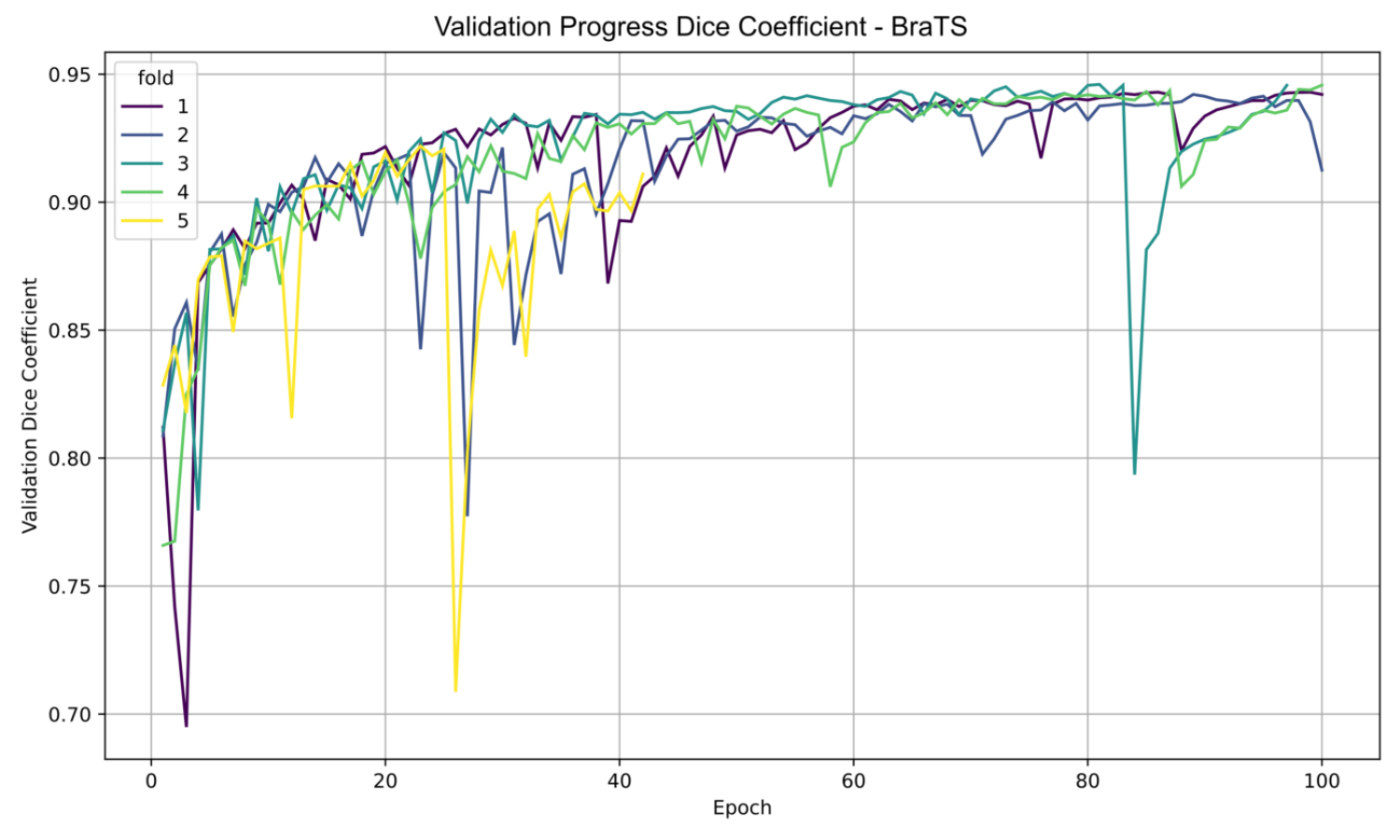}
        \caption{Validation Dice Score (BraTS)}
        \label{fig:curve_f}
    \end{subfigure}
    \caption{Per-fold learning curves. Left column: LGG dataset. Right column: BraTS 2020 dataset. Rows show, in order, training Focal Tversky loss, validation Focal Tversky loss, and validation Dice score. Each panel plots all five cross-validation folds.}
    \label{fig:curves_combined}
\end{figure}

\noindent The per-fold learning curves in Figure~\ref{fig:curves_combined} characterise the optimisation behaviour of the five fold models. Training Focal Tversky loss (Figures~\ref{fig:curve_a},~\ref{fig:curve_b}) and validation Focal Tversky loss (Figures~\ref{fig:curve_c},~\ref{fig:curve_d}) trend downward throughout training on both datasets. The curves are not monotone, and markedly so on BraTS 2020, where individual folds show repeated transient spikes, each followed within a few epochs by a return to the prevailing downward trend; one fold ends on such a spike at the final epoch. What matters for generalization is that neither validation-loss panel shows a sustained upward drift as training proceeds, since a sustained upturn is the signature we would expect if the model were fitting noise specific to the training folds. Validation Dice (Figures~\ref{fig:curve_e},~\ref{fig:curve_f}) rises steeply and then plateaus; the spread between folds is wide over the first half of training, but the five folds converge to a narrow band by the end, indicating that the final result does not depend strongly on which partition a model was trained on. Early stopping and learning-rate reduction were both driven by validation loss and terminate training at this plateau. We note that these curves describe optimisation behaviour on the validation folds during training; they are not the hold-out test results, which are reported separately in Table~\ref{tab:test_results}.

\subsection{Benchmarking and Computational Considerations}

\noindent To situate our result within the literature, Table~\ref{tab:performance_compare} lists reported whole-tumor Dice scores for several established segmentation methods on BraTS data. We emphasise that these entries are not measured under a common protocol, and we therefore make no claim of outperforming them. Our 95.10\% is an aggregate pixel-level $F_1$ over tumor-positive 2D FLAIR slices on a partition we constructed ourselves; the literature entries are per-case means over full 3D multimodal volumes, computed by the challenge organisers on withheld annotations. Because pooled-pixel aggregation is weighted by lesion size, and because our evaluation excludes the tumor-free slices on which false positives are most costly, our figure should be read as an upper bound relative to those numbers rather than as a like-for-like improvement.

\noindent We draw attention to this explicitly because the arithmetic invites the wrong conclusion. Our 95.10\% is numerically higher than the 92.94\% whole-tumor Dice with which the winning entry of the BraTS 2022 challenge~\cite{zeineldin2022multimodal} topped that year's testing leaderboard, and higher than Swin UNETR's 92.6\% on the BraTS 2021 validation set~\cite{hatamizadeh2021swin}. We do not believe our model is better than either, and we would regard any such reading as a misuse of this table. The gap is far more plausibly attributable to the differences in aggregation, dimensionality, modality count, and slice selection catalogued in the note above than to a genuine advance in segmentation quality. What the table does support is a narrower and, we believe, more useful claim: a 2D, single-modality, ImageNet-pretrained CNN with decoder-side channel attention operates in the same broad performance regime as substantially heavier 3D and transformer-based architectures, at a small fraction of their inference cost (Table~\ref{tab:computational_cost}).

\noindent One prior result on our second dataset warrants direct comment. Ghosh et al.~\cite{Ghosh2021} evaluate a VGG-16-based U-Net on the same TCGA-LGG collection used in our LGG experiments and report a pixel accuracy of 0.9975, against the 0.9930 obtained by our ensemble. We do not read this as a meaningful ranking in either direction. At the roughly 3\% tumor-pixel prevalence of this dataset, a predictor that labels every pixel as background already achieves approximately 97\% accuracy, so differences of a few tenths of a percentage point in this metric carry very little information about segmentation quality. We raise the comparison because it is the closest prior work to our LGG setting---same collection, same 2D formulation, same backbone family---and because it illustrates precisely why we treat overlap measures rather than pixel accuracy as the operative metric for this task.

\begin{table}[H]
    \centering
    \caption{Reported Dice scores for brain tumor segmentation methods on BraTS data. Evaluation protocols differ substantially across rows; see the note below.}
    \label{tab:performance_compare}
    \vspace{5pt}
    \renewcommand{\arraystretch}{1.35}
    \resizebox{\textwidth}{!}{%
    \begin{tabular}{|p{4.4cm}|c|c|c|c|p{4.6cm}|c|}
        \hline
        \textbf{Method} & \textbf{Dataset} & \textbf{Dim.} & \textbf{Modalities} & \textbf{Region} & \textbf{Evaluation protocol} & \textbf{Dice} \\
        \hline
        3D U-Net, multi-task learning~\cite{10.1007/978-3-030-46640-4_31} & BraTS 2019 & 3D & 4 & WT & Per-case mean, official validation set & 89.0\% \\
        \hline
        Swin UNETR~\cite{hatamizadeh2021swin} & BraTS 2021 & 3D & 4 & WT & Per-case mean, official validation set & 92.6\% \\
        \hline
        Multimodal CNN ensemble~\cite{zeineldin2022multimodal} & BraTS 2022 & 3D & 4 & WT & Per-case mean, official testing set (challenge winner) & 92.94\% \\
        \hline
        \textbf{Proposed VGG16-MCA UNet} & \makecell[c]{LGG (TCGA)\\BraTS 2020} & 2D & 1 (FLAIR) & WT & Pooled-pixel $F_1$, tumor-positive slices, own 85/15 split & \makecell[c]{88.32\%\\95.10\%} \\
        \hline
    \end{tabular}%
    }
\end{table}

\begin{flushleft}
    \footnotesize{\textbf{Note.} The rows in this table are not directly comparable and are not intended to establish a ranking. They differ in dataset year, input dimensionality (2D slices versus 3D volumes), number of input modalities, whether tumor-free slices are included in the evaluation, and---most consequentially---in how the Dice coefficient is aggregated. The three literature entries report a mean over patient cases on an official challenge evaluation set, computed by the organisers on withheld annotations. Our entry reports a single $F_1$ computed over all test pixels pooled into one confusion matrix, on tumor-positive slices only, using a train/test partition we constructed ourselves from the BraTS 2020 training release. Pooled-pixel aggregation is weighted by lesion size and is systematically higher than a per-case mean evaluated on the same predictions. The table is provided to situate our result within the literature, not to claim superiority over it.}
\end{flushleft}

\noindent Apart from segmentation accuracy, the real value of a model also lies in its computational efficiency. We compared the inference time and model complexity of our new model with some other UNet models, and the results are shown in Table~\ref{tab:computational_cost}. Our VGG16-MCA UNet has an average inference time of 66.32 ms. This is only an increase of around 8 ms from the baseline UNet with a VGG16 encoder (58.52 ms), showing that our domain-specific Multi-Channel Attention module incurs a negligible computational cost.

\noindent Our model also compares favourably on cost against configurations built on deeper backbones: the ResNet50 encoder U-Net measured over three times longer (222.02 ms) in the same setting, without a corresponding gain in segmentation accuracy. Subject to the measurement caveats noted below Table~\ref{tab:computational_cost}, this suggests that the chosen encoder--decoder configuration occupies a favourable region of the accuracy--cost trade-off for single-image inference on commodity hardware, which is the setting we target.

\begin{table}[H]
\centering
\caption{Computational Cost Comparison Of Different Segmentation Model}
\label{tab:computational_cost}
\resizebox{\textwidth}{!}{%
\begin{tabular}{|l|r|r|r|}
\hline
\textbf{Model} & \textbf{FLOPs} & \textbf{Params} & \textbf{Avg Inference Time (ms)} \\ \hline
VGG16\_Backbone\_Only     & 40,114,847,744   & 14,714,688  & 18.74 \\ \hline
VGG19\_Backbone\_Only     & 50,988,187,648   & 20,024,384  & 21.14 \\ \hline
ResNet50\_Backbone\_Only  & 10,123,832,960   & 23,587,712  & 178.33 \\ \hline
ResNet101\_Backbone\_Only & 19,845,730,944   & 42,658,176  & 445.51 \\ \hline
U\_Net\_from\_Scratch    & 96,446,548,480   & 31,055,297  & 58.51 \\ \hline
UNet\_PlusPlus       & 69,076,604,308   & 9,170,625  & 139.73 \\ \hline
U\_Net\_VGG16            & 101,918,387,456  & 25,862,337  & 58.52 \\ \hline
U\_Net\_VGG19            & 112,791,727,360  & 31,172,033  & 47.44 \\ \hline
U\_Net\_ResNet50         & 29,601,291,136   & 40,977,665  & 222.02 \\ \hline
U\_Net\_ResNet101        & 39,323,189,120   & 60,048,129  & 400.62 \\ \hline
Proposed Model      & 101,952,632,896  & 27,257,857  & 66.32 \\ \hline
\end{tabular}%
}
\end{table}

\begin{flushleft}
    \footnotesize{*Inference time measured for a single $256\times256$ image on an NVIDIA RTX 2060 GPU with 6\,GB VRAM. These are single-run, batch-size-one measurements taken through the training framework, and therefore include per-call framework and kernel-launch overhead, which is not negligible at this input size and does not scale with model FLOPs. The figures should be read as indicative of the relative cost of each architecture in a comparable single-image deployment setting, rather than as precise throughput benchmarks; in particular, the ordering of models whose measured times differ by less than roughly 20\% should not be regarded as significant.}
\end{flushleft}
\subsection{Qualitative Analysis}
\begin{figure}[H]
    \centering
    \begin{subfigure}{0.48\textwidth}
        \includegraphics[width=\linewidth]{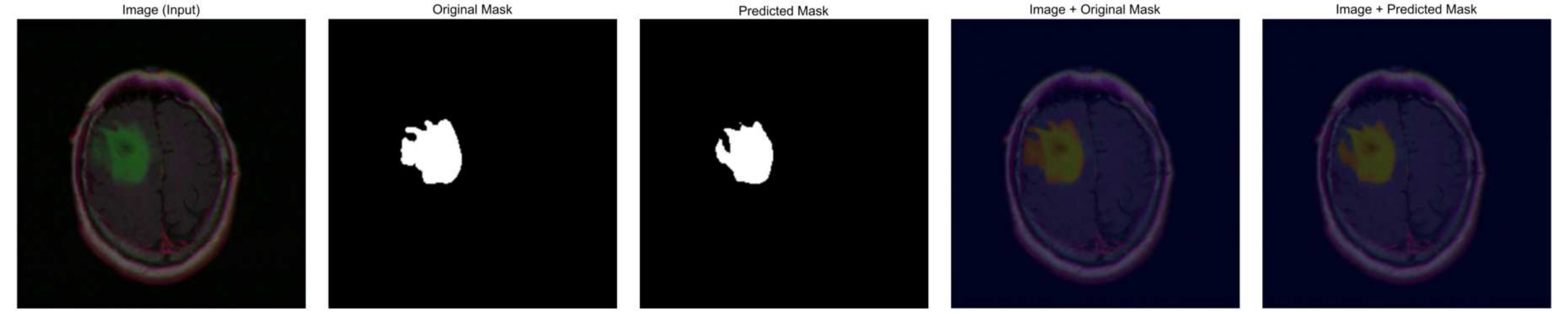}
        \caption{LGG Sample 1}
        \label{fig:lgg_pred_1}
    \end{subfigure}
    \hfill
    \begin{subfigure}{0.48\textwidth}
        \includegraphics[width=\linewidth]{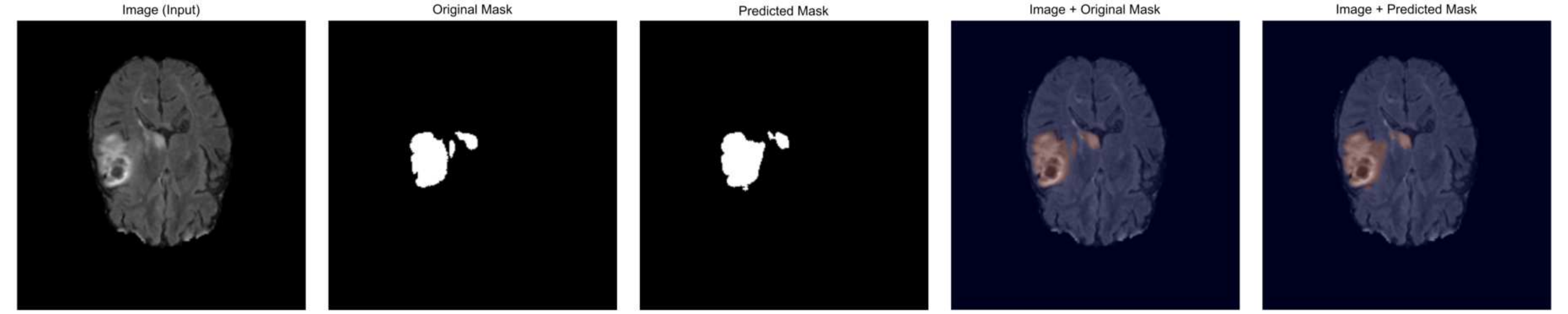}
        \caption{BraTS Sample 1}
        \label{fig:brats_pred_1}
    \end{subfigure}
    \vspace{0.5cm}
    \begin{subfigure}{0.48\textwidth}
        \includegraphics[width=\linewidth]{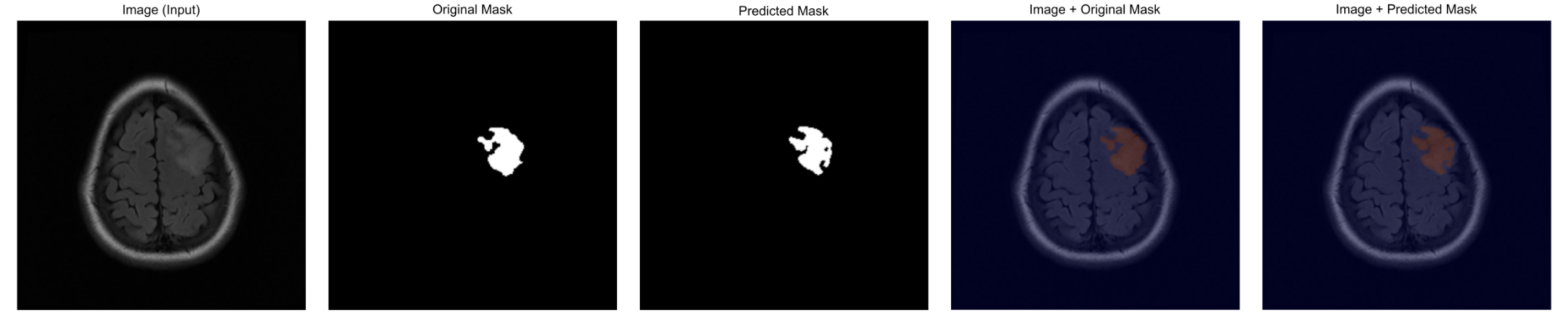}
        \caption{LGG Sample 2}
        \label{fig:lgg_pred_2}
    \end{subfigure}
    \hfill
    \begin{subfigure}{0.48\textwidth}
        \includegraphics[width=\linewidth]{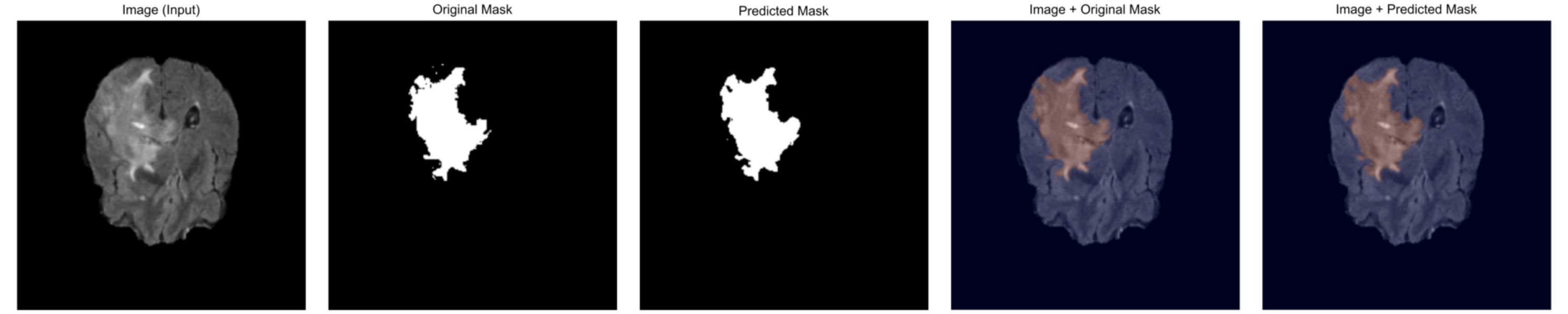}
        \caption{BraTS Sample 2}
        \label{fig:brats_pred_2}
    \end{subfigure}
    \vspace{0.5cm}
    \begin{subfigure}{0.48\textwidth}
        \includegraphics[width=\linewidth]{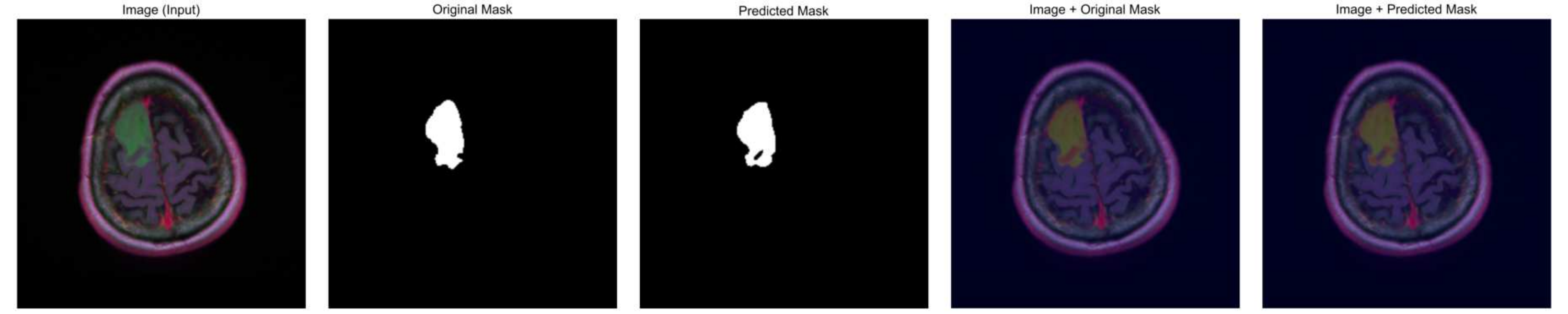}
        \caption{LGG Sample 3}
        \label{fig:lgg_pred_3}
    \end{subfigure}
    \hfill
    \begin{subfigure}{0.48\textwidth}
        \includegraphics[width=\linewidth]{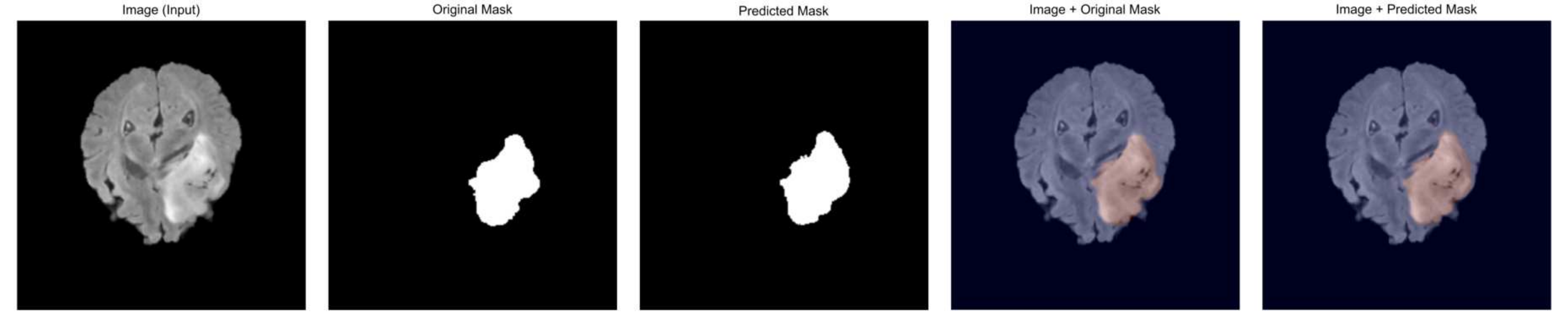}
        \caption{BraTS Sample 3}
        \label{fig:brats_pred_3}
    \end{subfigure}
    \caption{Qualitative segmentation outcome of the proposed VGG16-MCA UNet ensemble model on sample test images from both datasets. Left column shows LGG dataset samples, right column shows BraTS 2020 dataset samples. The input image, ground truth mask, and the segmentation mask predicted by our model are displayed in each subfigure. The model demonstrates excellent boundary delineation and tumor region identification across different tumor morphologies and sizes.}
    \label{fig:model_pred}
\end{figure}

\noindent Figure~\ref{fig:model_pred} shows a qualitative comparison of segmentation masks produced by our ensemble model and ground truth for sample test images of both the BraTS 2020 and LGG datasets. The model accurately determines precise tumor boundaries, even in cases involving complex shapes and tumors of different sizes.

\noindent Our qualitative results show some of the key strengths of our suggested model. For the LGG dataset examples (Figures~\ref{fig:lgg_pred_1}-\ref{fig:lgg_pred_3}), the model is able to pick up the subtle edges of lower-grade gliomas, which are typically less aggressive in their growth and more diffuse in their borders. The segmentation masks of the model have excellent correlation with ground truth, particularly over regions where tumor tissue gradually transitions to adjacent healthy brain tissue. The model is also observed to be robust to vastly different tumor sizes and shapes, from small focal lesions to large infiltrating tumors.
\noindent For BraTS 2020 data samples (Figures~\ref{fig:brats_pred_1}-\ref{fig:brats_pred_3}), which include low-grade and high-grade gliomas, the model demonstrates strong performance on a range of tumor attributes. The predictions mark tumor regions with a range of levels of contrast enhancement and edema with clear separation, which is a demonstration of the capability of the model in dealing with heterogeneity in brain tumor imaging. Segmentation curves tightly follow ground truth curves, a confirmation to the capability of the MCA mechanism to concentrate on significant tumor features and discard background noise. The model demonstrates stable performance on a range of tumor size and shape, from small focal lesions to large infiltrating tumors. This variety is of particular importance to clinical application where tumor appearances can be quite variable from patient to patient.
\FloatBarrier
\section{Limitations}
\label{sec:limitations}
\noindent The scope of this study is deliberately narrow and should be read as such. We segment a single binary target---the whole tumor---from 2D axial FLAIR slices, and we train and evaluate only on slices that contain tumor. We therefore do not address the tumor core and enhancing tumor sub-regions that the BraTS protocol scores, do not exploit the T1, T1ce, and T2 volumes available in BraTS 2020, do not model inter-slice context, and do not measure false-positive behaviour on tumor-free slices---which is where a system deployed over whole volumes would accumulate much of its error. All overlap figures we report are pooled-pixel statistics rather than per-case means, for the reasons set out in Section~\ref{sec:results}; they are a valid characterisation of the model but are not interchangeable with challenge-protocol Dice. Both test sets are internal to their respective public collections, so the generalization evidence here spans two datasets rather than two institutions. The two collections are moreover not independent cohorts: the BraTS training archive incorporates TCGA-LGG cases drawn from the same pool as the LGG MRI Segmentation dataset, so an unknown number of patients may be common to both. We have not computed this intersection, and the agreement between our two sets of results should be discounted accordingly. A more consequential limitation concerns how the data were partitioned. We have verified, against both our training code and the recorded split files, that the 85/15 hold-out split and the five cross-validation folds were drawn over individual slices rather than over patients. Consequently all 110 LGG patients and all 368 BraTS 2020 cases contribute slices to both the training folds and the hold-out test set. Because consecutive slices through one volume are highly correlated, the model has seen immediate neighbours of every test slice during training, and the figures reported here therefore measure interpolation within known patients rather than generalization to unseen ones. Published estimates of this effect in brain MRI are large, and on the LGG collection specifically have been put at roughly ten to twenty Dice points. Our results should accordingly be read as an upper bound, and should not be set against results obtained under patient-level protocols, which includes every entry in Table~\ref{tab:performance_compare}. We report them because they are what our pipeline computed and because the protocol that produced them is stated in full and the split records themselves are released, not because we regard them as an estimate of performance on new patients. Finally, the latency figures in Table~\ref{tab:computational_cost} are single-run, batch-size-one measurements taken on one consumer GPU and should be treated as indicative of relative cost rather than as precise throughput benchmarks.

\section{Conclusion and Future Work}
\noindent In this work we presented the VGG16-MCA UNet, a hybrid architecture for whole-tumor segmentation in 2D FLAIR MRI. The model combines a pre-trained VGG16 encoder with a decoder in which a Multi-Channel Attention module recalibrates features after each skip-connection fusion, trained under the Focal Tversky loss with 5-fold cross-validation and ensembling. On our held-out test splits it reached an aggregate pixel-level Dice of 95.10\% on BraTS 2020 and 88.32\% on LGG, at an inference cost of 66.32 ms per slice on a single 6\,GB consumer GPU---approximately 8 ms more than the same encoder--decoder without the attention modules. These figures were obtained under slice-level partitioning, and as set out in Section~\ref{sec:limitations} they characterise the pipeline's behaviour within patients it has already seen rather than its performance on new ones. We therefore present this work as a fully specified and reproducible 2D FLAIR baseline, not as evidence of clinical readiness.

\noindent Several directions follow directly. The first and most important is to repeat this evaluation with the hold-out split and the cross-validation folds constrained at the patient level, so that no patient contributes slices to more than one partition, and to report the difference between the two protocols on identical data. That comparison requires no new data and no change of architecture, and it would place a quantitative bound on how much slice-level partitioning inflates results on these two widely used collections---a question the surrounding literature has largely left open. Beyond that, we would investigate quantization and pruning to reduce inference cost for deployment in resource-constrained settings, and extend the evaluation to 3D volumetric data. This may provide useful insight and enhance the performance of segmentation. Lastly, the ultimate goal is to apply this in actual clinical environments; thus, we will strive to test our model in upcoming clinical studies with medical organizations to determine how it impacts patient care in practice.
\section*{Data Availability}
\noindent Both datasets used in this study are publicly available, and no new data were generated. The LGG MRI Segmentation dataset~\cite{buda2019association, dataset} is distributed through Kaggle at \url{https://www.kaggle.com/datasets/mateuszbuda/lgg-mri-segmentation} and derives from The Cancer Genome Atlas lower-grade glioma collection hosted by The Cancer Imaging Archive. The BraTS 2020 dataset~\cite{menze2014multimodal, bakas2018identifying} is available from the Center for Biomedical Image Computing and Analytics at \url{https://www.med.upenn.edu/cbica/brats2020/data.html}; access requires registration through the CBICA Image Processing Portal. We used the BraTS 2020 training partition only, since ground-truth annotations for the official validation and test partitions are not publicly released.

\section*{Code Availability}
\noindent The model definition, the training and evaluation scripts, the per-fold training histories, and the exact partition indices used to produce every result reported in this paper are available at \url{https://github.com/ShubhamGajjar/vgg16-mca-unet}. We release the partition files specifically so that the evaluation protocol described in Section~\ref{sec:limitations}, including its slice-level granularity, can be verified directly rather than taken on trust. Trained weights are available from the corresponding author on request.

\bibliographystyle{unsrtnat}
\bibliography{references}

\end{document}